\documentclass[Times,4pt,aps,pra,twocolumn,showpacs,amsmath,amssymb,floatfix,footinbib,superscriptaddress,times]{revtex4-1}
\usepackage{graphics,graphicx,epsfig,ulem,epstopdf,bm,longtable,url,datetime}
\usepackage[colorlinks=true,linkcolor=red,urlcolor=red,citecolor=red]{hyperref}
\usepackage{amsmath,amssymb,latexsym,float,amsfonts,subfigure,hyperref,color}
\usepackage[ansinew]{inputenc}
\usepackage[usenames,dvipsnames]{pstricks}
\usepackage[mathlines]{lineno}
\def\footnoterule{\kern -1mm \hrule width 5.8cm \kern 2.2mm}%
\usepackage{tikz,xcolor,hyperref}% Make Orcid icon
\definecolor{lime}{HTML}{A6CE39}
\DeclareRobustCommand{\orcidicon}{%
    \begin{tikzpicture}
    \draw[lime, fill=lime] (0,0)
    circle [radius=0.16]
    node[white] {{\fontfamily{qag}\selectfont \tiny ID}};\draw[white, fill=white] (-0.0625,0.095)
    circle [radius=0.007];
    \end{tikzpicture}
    \hspace{-2mm}}
\foreach \x in {A, ..., Z}
{\expandafter\xdef\csname orcid\x\endcsname{\noexpand\href{https://orcid.org/\csname orcidauthor\x\endcsname}{\noexpand\orcidicon}}}

\begin{document}
\title{  Left-handedness with three zero-absorption windows tuned by the incoherent pumping field and inter-dot tunnelings in a GaAs/AlGaAs triple quantum dots system }
\thanks{ Supported by National Natural Science Foundation of China (NSFC) (Grant No.61205205)
and the Foundation for Personnel training projects of Yunnan Province (grant No.KKSY201207068) of China.}

\author{Shun-Cai Zhao\orcidA{}}%
\email[Corresponding author: ]{zhaosc@kmust.edu.cn.}
\affiliation{Faculty of science, Kunming University of Science and Technology, Kunming, 650093, PR China}
\affiliation{Center for Quantum Materials and Computational Condensed Matter Physics, Faculty of Science, Kunming University of Science and Technology, Kunming, 650500, PR China}

\author{Shuang-Ying Zhang}
\affiliation{Faculty of science, Kunming University of Science and Technology, Kunming, 650093, PR China}
\affiliation{Center for Quantum Materials and Computational Condensed Matter Physics, Faculty of Science, Kunming University of Science and Technology, Kunming, 650500, PR China}

\author{Qi-Xuan Wu}
\affiliation{ College English department, Kunming University of Science and Technology, Kunming, 650500, PR China}

\author{Jing Jia}
\affiliation{Faculty of science, Kunming University of Science and Technology, Kunming, 650093, PR China}
\affiliation{Center for Quantum Materials and Computational Condensed Matter Physics, Faculty of Science, Kunming University of Science and Technology, Kunming, 650500, PR China}
\begin{abstract}
Left-handedness with three zero-absorption windows is achieved in a triple-quantum dot (TQD) system. With the typical parameters of a GaAs/AlGaAs heterostructure, the simultaneous negative relative electric permittivity and magnetic permeability are obtained by the adjustable incoherent pumping field and two inter-dot tunnelings. What's more, three zero-absorption windows in the left-handedness frequency bands are observed in the TQD system. The left-handedness with zero-absorption in solid state heterostructure may solve not only the challenge in the photonic resonant scheme for left-handed materials (LHMs) but also the application limitation of the negative refractive materials with large amount of absorption.
\end{abstract}

\maketitle
\section{Introduction}

Recently, many experimental and theoretical investigations have been devoted to understanding
and construction of micro-structured materials which enjoy exotic optical properties, such as
left-handedness\cite{1}. In such a material characterized by negative electric and magnetic permittivity
(left-handed materials, LHMs), many surprising phenomena occur: amplification of evanescent waves\cite{2},
sub-wavelength focusing\cite{3}, reversals of both Snell's law and Doppler shift\cite{1}, etc. Artificial structures\cite{4,5,6},
photonic crystals\cite{7,8}, chiral materials\cite{9,10,11} and photonic resonant media\cite{12,13,14} were brought birth to LHMs. However,
in the photonic resonant media scheme large external magnetic fields were required  because of the
major challenge of two separated levels with a non-vanishing magnetic dipole matrix element at
optical frequencies\cite{8}. So the excitonic energy levels in solid state heterostructures maybe the candidate
for the challenge.

With the flaw of photonic resonant media scheme in mind, the triple quantum dot(TQD) arrays fabricated in a GaAs/AlGaAs heterostructure\cite{15} may be an ideal candidate because the TQD possesses the atom vapors' features in solid-state structures and provides a fully
tunable platform via the coherent coupling of quantum states. Recently, the TQD system has attracted the intense investigation\cite{16,17,18,19,20,21,22,23,24} because of its potential applications in quantum information. The corresponding investigations include the charge and spin transport properties\cite{18}, the manipulation of three-electron spin states\cite{19}, Spin-orbit effects\cite{20}, a resonant exchange qubit\cite{21} and spin blockade turning to bipolar\cite{22,23}. The typical TQD can be grown by molecular epitaxy on [0 0 1] GaAs substrate with the typical heights span of 1.5-2.5 nanometer when the InAs is imbedded into GaAs/Al$_{0.45}$Ga$_{0.55}$As layer\cite{24}. The volumes of the triple dots can be modeled as three spheres with 3.5 nm high and 35 nm in diameter for QD1, 2.5 nm high and
25 nm in diameter for QD2 and QD3, respectively\cite{25}, as shown in Fig.1(a).

Here we theoretically investigate the left-handedness and absorption property in this TQD. It finds that Left-handedness
with three zero-absorption windows tuned by the incoherent pumping field and two inter-dot tunnelings of this TQD is achieved,
as is important due to overcome the defect in the photonic resonan scheme LHMS and the application limitation of negative refractive materials with large amount of absorption.

\section{Model and equation}

The cross section of the TQD nanostructure under consideration is shown in Fig.1(a), and its corresponding energy levels is
depicted in Fig.1(b). The energy levels of the TQD molecule are consists of three dots,
which are coupled by the electron tunneling and form a four atomic levels. Levels
$|1\rangle$ and $|2\rangle$ are the lower and upper conducting band levels of QD1,
respectively. Levels $|3\rangle$ and $|4\rangle$ are the excited conducting level
of QD2 and QD3, respectively. The energy differences of the three excited conducting levels
between the lower level are assumed to be large, so their corresponding tunneling couplings are ignored.
However, the tunnel barrier can be manipulated by the positioned gate electrode between the neighboring QDs\cite{26}.
Thus, levels $|3\rangle$ and $|4\rangle$ can get closer to level $|2\rangle$ via the gate voltage. An
incoherent pump field with pumping rate denoted by $\Gamma$ and a weak probe field with Rabi frequency $\Omega_{p}$ and frequency $\omega_{p}$
are coupling the transitions $|1\rangle$ and $|2\rangle$. The electric and magnetic components (corresponding Rabi frequency $\Omega_{e}$=$\rho_{21}E_{p}/\hbar$ and $\Omega_{b}$=$m_{31}B_{p}/\hbar$ ) of the
probe field interact with the transitions $|1\rangle$ and$|2\rangle$ as well as $|1\rangle$ and$|3\rangle$, respectively.
Under the coupling of the probe field with the QD1, the electron can be excited from the bands
$|1\rangle$ to $|2\rangle$, then be transferred to QD2 and QD3 (the levels $|3\rangle$ and $|4\rangle$)
via the tunneling $T_{a}$ and $T_{b}$. $T_{a}$ and $T_{b}$ are the electron tunneling matrix elements
between the triple QDs. The probability of electron tunneling from the first QD1 in state $\Psi$
with energy $E_{\Psi}$ to the second QD2 in state $\Phi$ with energy $E_{\Phi}$ can describe by Fermi's
golden rule\cite{27}: $ p=\frac{2\pi}{\hbar}|T_{e}|^{2}\delta(E_{\Psi}-E_{\Phi})$. So the tunneling matrix element
in the barrier region between the QDs can be integrated over a surface:

\begin{figure}[htp]
\center
\includegraphics[totalheight=0.8 in]{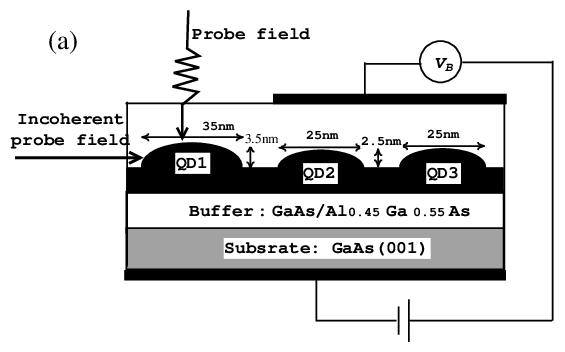 }
\includegraphics[totalheight=0.8 in]{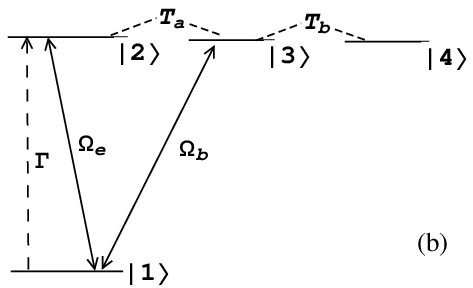 }
\caption{(a) Cross section of the InAs TQD imbedded into a GaAs/Al$_{0.45}$Ga$_{0.55}$As layer with the probe and incoherent pumping fields. (b) The quantized energy levels for the TQD interacting with the probe and incoherent pumping fields. The electric and magnetic components of the probe field are coupled to the level pairs $|1\rangle$-$|2\rangle$ and $|1\rangle$-$|3\rangle$, respectively.}
\end{figure}\label{Fig.1}

\begin{equation}
T_{e}=\frac{\hbar}{2m}\int_{{z=z_{0}}}({\Phi^{*}\frac{\partial\Psi}{\partial z}-\Psi^{*}\frac{\partial\Phi^{*}}{\partial z}})ds
\end{equation}

in which m is the electron's effective mass and $z_{0}$ is in the barrier. The tunneling matrix element $T_{e}$ can be tuned
by modifying the bias\cite{28} applied to the QDs molecule because of the relation

\begin{equation}
I=\frac{4e\pi}{\hbar}\int^{eV}_{{0}}\varrho_{1}(E_{F}-eV+\varepsilon)\varrho_{2}(E_{F}+\varepsilon)|T_{e}|^{2}d\varepsilon
\end{equation}

where I is the current which is proportional to the local state density of each QD ($\varrho_{1}$, $\varrho_{2}$)
with the Fermi energy ($E_{F}$). Under the rotating wave approximation, we write the total Hamiltonian describing
the interaction of the probe and incoherent pumping fields with the TQD with neglecting from the hole tunneling as follows

\begin{eqnarray}
&H= &\sum^{4}_{i=1}E_{i}|i\rangle\langle i|+[(\Omega_{e}e^{-i\omega_{p}t}|1\rangle\langle 2|+\Omega_{b}e^{-i\omega_{p}t}|1\rangle\langle 3| \nonumber \\
&&+\varepsilon D|1\rangle\langle 2|+T_{a}|2\rangle\langle 3|+T_{b}|3\rangle\langle 4|)+H.C. ]
\end{eqnarray}

where $E_{i}$=$\hbar\omega_{i}$ is the energy of level $|i\rangle$, and $\varepsilon$ is the electrical
amplitude of the incoherent pumping field with D being the dipole moment corresponding to the transition
from $|1\rangle$ to $|2\rangle$. Using the density-matrix approach, the time-evolution of the
TQD system can be obtained as

\begin{eqnarray}
&\dot{\rho}_{11}=&(\gamma_{21}+\Gamma)\rho_{22}+\gamma_{31}\rho_{33}+\gamma_{41}\rho_{44}+i(\Omega_{e}\rho_{21}-\Omega^{*}_{e}\nonumber\\
                &&\rho_{12})-\Gamma\rho_{11}, \nonumber \\
&\dot{\rho}_{22}=&-(\gamma_{21}+\Gamma)\rho_{22}-i(\Omega_{e}\rho_{21}-\Omega^{*}_{e}\rho_{12})+i(T^{*}_{a}\rho_{32}-T_{a}\nonumber \\
                &&\rho_{23})+\Gamma\rho_{11}, \nonumber \\
&\dot{\rho}_{33}=&-\gamma_{31}\rho_{33}+i(T_{a}\rho_{23}-T^{*}_{a}\rho_{32})-i(T_{b}\rho_{34}-T^{*}_{b}\rho_{43}), \nonumber \\
&\dot{\rho}_{44}=&-\gamma_{41}\rho_{44}+i(T_{b}\rho_{34}-T^{*}_{b}\rho_{43}), \nonumber \\
&\dot{\rho }_{12}=&(i\Delta_{p}-\Gamma_{21}-\Gamma)\rho_{12}-i\Omega_{e}(\rho_{11}-\rho_{22})-i T_{a}\rho_{13},\nonumber \\
&\dot{\rho }_{13}=&[i(\Delta_{p}+\omega_{12})-\Gamma_{31}-\Gamma/2]\rho_{13}+i\Omega_{b}\rho_{23}-i (T^{*}_{a}\rho_{12} \nonumber \\
                 &&+T_{b}\rho_{14}), \label{eq4}\\
&\dot{\rho }_{14}=&[i(\Delta_{p}+\omega_{12}+\omega_{23})-\Gamma_{41}-\Gamma/2]\rho_{14}+i\Omega_{e}\rho_{24}\nonumber \\
                 &&-i T^{*}_{b}\rho_{13}, \nonumber \\
&\dot{\rho }_{32}=&-(i\omega_{23}+\Gamma_{12}+\Gamma/2)\rho_{32}-i T_{a}(\rho_{33}-\rho_{22})-i\Omega_{b}\rho_{31}\nonumber \\
                 &&+i T^{*}_{b}\rho_{42},\nonumber \\
&\dot{\rho }_{42}=&-[i(\omega_{23}+\omega_{34})+\Gamma_{42}+\Gamma/2]\rho_{42}-i(T_{a}\rho_{43}-T_{b}\rho_{32})\nonumber \\
                 &&-i\Omega_{e}\rho_{41},\nonumber \\
&\dot{\rho}_{43}=&-(i\omega_{34}+\Gamma_{43})\rho_{43}-i T_{b}(\rho_{44}-\rho_{33})-i T^{*}_{a}\rho_{42},\nonumber \\
&1=&\rho_{11}+\rho_{22}+\rho_{33}+\rho_{44}\nonumber
\end{eqnarray}

where $\omega_{ij}$=$\omega_{i1}$-$\omega_{j1}$,$ i,j$=2,3,4. $\Delta_{p}$=$\omega_{21}$-$\omega_{p}$
The total decay rates of the TQD are added phenomenologically in above density matrix equations (4), which include the
spontaneous emission rates and dephasing rates. The spontaneous emission rates for sub-band $|i\rangle$ because of the longitudinal optical (LO) phonon emission events\cite{29} are denoted by $\gamma_{i1}$. The total decay rates $\Gamma_{ij}$ ($i\neq j$) are given
by $\Gamma_{i1}$=$ \gamma_{i1}/2$ + $\gamma^{dph}_{i1}$, $\Gamma_{ij}$=$ (\gamma_{i1}+\gamma_{j1})/2$ + $\gamma^{dph}_{ij}$, $i,j$=2,3,4 and $i\neq j$, in which $\gamma^{dph}_{ij}$ is the dominant mechanism in a semiconductor solid-state system and determined by electron-electron,
interface roughness and phonon scattering processes.

In the TQD molecule the classical electric polarizability defined by its Fourier transform $\vec{p}_{e}(\omega_{p})$ $=\epsilon_{0}$$\alpha_{e}(\omega_{p})\vec{E}(\omega_{p})$ is a rank 2 tensor and is related to the mean value of the
atomic electric-dipole moment operators via the definition $\vec{P}_{e}$=Tr$\{{\hat{\rho}\vec{d}}\}$=$\rho_{12}d_{21}$+c.c., where Tr
stands for trace. In the following, the polarizability at the frequency $\omega_{P}$ of the incident probe field
$\vec{E}_{p}$ is considered, and we pay close attention to the $\omega_{P}$ dependence $\alpha_{e}(\omega_{P})\equiv\alpha_{e}$.
When $\vec{E}_{p}$ parallels to the atomic dipole $\vec{d}_{21}$, $\alpha_{e}$ is a scalar and its expression is obtained:

\begin{equation}
\alpha_{e}=\frac{\vec{d}_{21}\rho_{12}}{\epsilon_{0}\vec{E}_{p}}=\frac{\mid
{d_{21}}\mid^{2} \rho_{12}}{\epsilon_{0}\hbar\Omega_{e}},
\end{equation}

Similarly, the classical magnetic polarizations of the TQD molecule $\vec{P}_{m}(\omega_{P})$=$\mu_{0}\alpha_{m}\vec{E}(\omega_{p})$, which
is related to the mean value of the dipole moment operator through $\vec{P}_{m}$=Tr$\{{\hat{\rho}\vec{\mu}}\}$=$\rho_{31}\mu_{13}$+c.c.. When the TQD molecule's magnetic dipole is perpendicular to the induced electric dipole, the magnetizability $\alpha_{m}$ is a scalar via the classical Maxwell's electromagnetic wave-vector relation. Then $\alpha_{m}$ is achieved as follows:
\begin{equation}
\alpha_{m}=\frac{\mu_{0}\vec{\mu}_{13}\rho_{31}}{\vec{B}_{p}}=\frac{\mu_{0}\mid\mu_{13}\mid^{2}\rho_{31}}{\hbar\Omega_{b}}.
\end{equation}
According to the Clausius-Mossotti relations, the relative permittivity and relative
permeability of the TQD system are expressed as\cite{30,31}
\begin{eqnarray}
\epsilon_{r}=\frac{1+\frac{2}{3}N\alpha_{e}}{1-\frac{1}{3}N\alpha_{e}},
\mu_{r}=\frac{1+\frac{2}{3}N\gamma_{m}}{1-\frac{1}{3}N\gamma_{m}}.
\end{eqnarray}

Then, the expressions for the electric permittivity and magnetic permeability of the TQD molecule are obtaned.
In the section that follows, the left-handedness and absorption property in the TQD are discussed numerically.
The simultaneously negative permittivity and permeability, negative refraction
with triple zero-absorptive windows will be observed in this TQD system.

\section{Results and discussions}

\begin{figure}[htp]
\center
\includegraphics[totalheight=0.8 in]{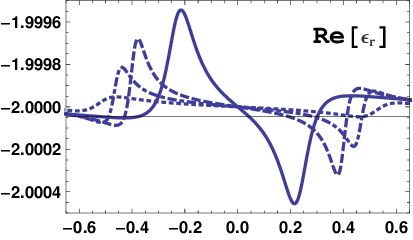 }\includegraphics[totalheight=0.8 in]{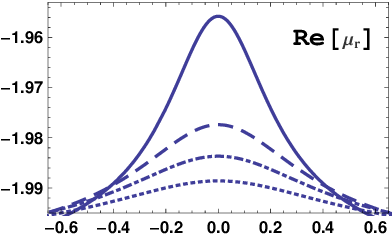  }
\hspace{0in}%
\includegraphics[totalheight=0.8 in]{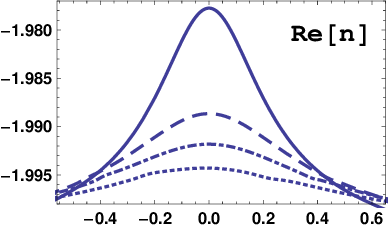 }\includegraphics[totalheight=0.8 in]{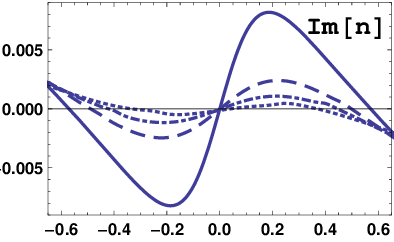 }
\caption{(Color online) $Re[\epsilon_{r}]$, $Re[\mu_{r}]$, $Re[n], $ $Im[n]$ versus the probe detuning $\frac{\Delta_{p}}{\Gamma_{10}}$. Left-handedness tuned by the incoherent pumping field $\Gamma$: $\Gamma$=$0$(dot curves), $\Gamma$=0.2 $\Gamma_{10}$(dash and dot curves), $\Gamma$=0.3 $\Gamma_{10}$ (dash curves), $\Gamma$=0.4 $\Gamma_{10}$(solid curves), where $\Omega_{p}$=0.05 $\Gamma_{10}$, $T_{a}$=0.25 $\Gamma_{10}$, $T_{b}$=0.60 $\Gamma_{10}$, $\omega_{12}$ =$\omega_{23}$=0. $\Gamma_{10}$=1.6 $\gamma$, $\gamma$=1 GHz.}
\end{figure}\label{Fig.2}

In this section we investigate left-handedness and absorption property of the TQD by using the numerical results. Several typical parameters of the GaAs/AlGaAs heterostructure are selected from the practical case: the sheet density of the TQD is taken as $3.7\times 10^{11}cm^{-2}$ \cite{32}, the electric transition dipole moment is chosen from the measured parameter: $ d_{12}=2.335\times1.602\times10^{-19}C m$\cite{33}, and the magnetic transition dipole moment is chosen from the typical parameter $ \mu_{23}=7.0\times10^{-23}Cm^{2}s^{-1}$\cite{34}. The other parameters are scaled by the typically slow dephasing rates $\Gamma_{10}$=1.6 $\gamma$, $\gamma$=1 GHz\cite{35}: $\Omega_{p}$=0.05 $\Gamma_{10}$, $\omega_{12}$ =$\omega_{23}$=0.

\begin{figure}[htp]
\center
\includegraphics[totalheight=0.8 in]{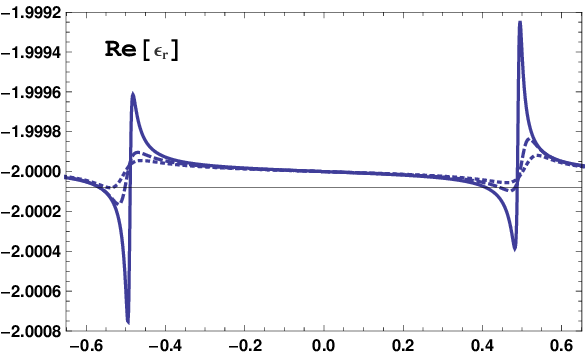 }\includegraphics[totalheight=0.8 in]{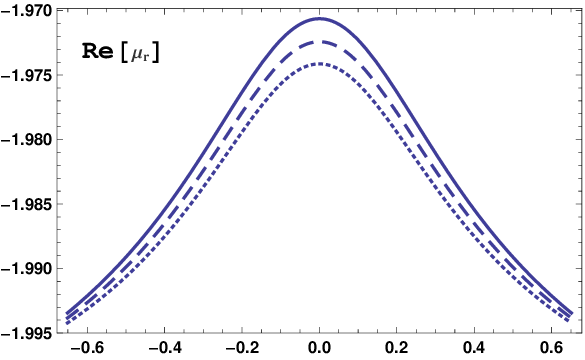  }
\hspace{0in}%
\includegraphics[totalheight=0.8 in]{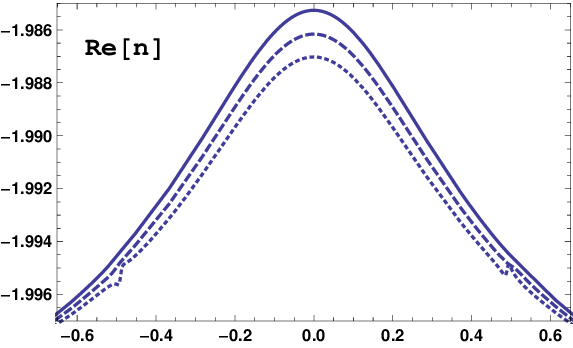 }\includegraphics[totalheight=0.8 in]{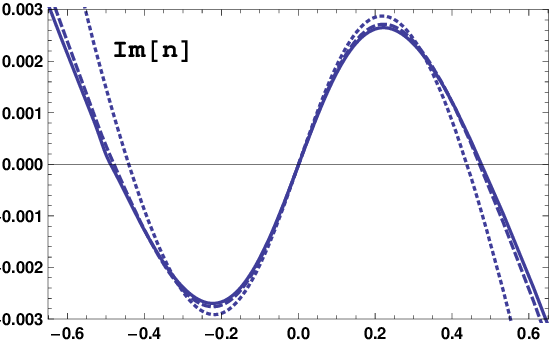 }
\caption{(Color online) $Re[\epsilon_{r}]$, $Re[\mu_{r}]$, $Re[n], $ $Im[n]$ versus the probe detuning $\frac{\Delta_{p}}{\Gamma_{10}}$. Left-handedness tuned by the tunneling $T_{a}$:  $T_{a}$=0.23 $\Gamma_{10}$(dot curves), $T_{a}$=0.235 $\Gamma_{10}$(dash curves), $T_{a}$=0.24 $\Gamma_{10}$(solid curves), $\Gamma$=0.2$\Gamma_{10}$, and the other parameters are the same to Fig.2.}
\end{figure}\label{Fig.3}

Fig.2 shows the calculated real parts of the relative electric permittivity$Re[\epsilon_{r}]$, magnetic permeability $Re[\mu_{r}]$ and relative refractive index $n$ of the TQD system as a function of probe frequency detuning $\Delta_{p}$, which are tuned by the incoherent pumping field. The absorption property of the TQD system can be described by the imaginary part of relative refractive index $Im[n]$. The left-handedness with simultaneous negative $Re[\epsilon_{r}]$, $Re[\mu_{r}]$ and $Re[n]$ is observed in Fig.2. It should be noted the increasing incoherent pumping field can enhance the negative response of the TQD system, which is shown by the dot curves($\Gamma$=$0$), dash and dot curves($\Gamma$=0.2 $\Gamma_{10}$), dash curves($\Gamma$=0.3 $\Gamma_{10}$) and solid curves($\Gamma$=0.4 $\Gamma_{10}$). This is the consequence of more electrons excited from the bands $|1\rangle$ to $|2\rangle$ by the incoherent pumping field. However, the triple transparency windows shown by the imaginary part of relative refractive index $Im[n]$ with two of them symmetrically distributing on both sides of the resonant position are remained in the left-handed frequency-band when the incoherent pumping field is tuned.

\begin{figure}[htp]
\center
\includegraphics[totalheight=0.8 in]{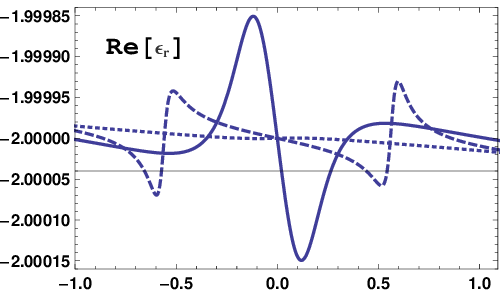 }\includegraphics[totalheight=0.8 in]{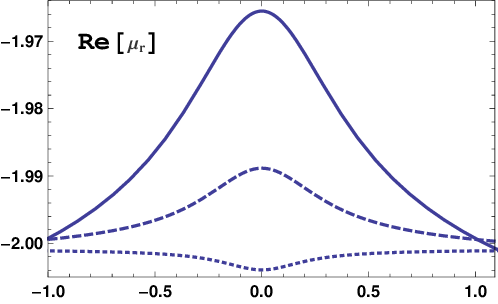  }
\hspace{0in}%
\includegraphics[totalheight=0.8 in]{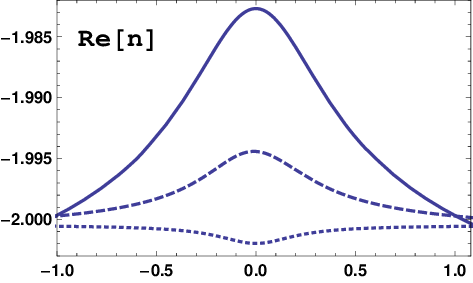 }\includegraphics[totalheight=0.82 in]{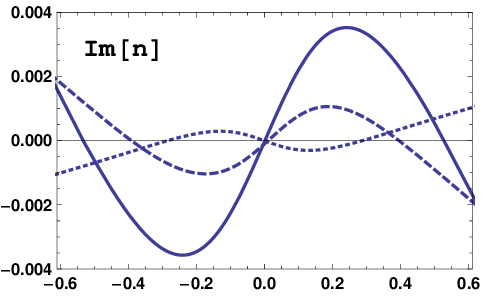 }
\caption{(Color online)(Color online) $Re[\epsilon_{r}]$, $Re[\mu_{r}]$, $Re[n], $ $Im[n]$ versus the probe detuning $\frac{\Delta_{p}}{\Gamma_{10}}$. Left-handedness tuned by the tunneling $T_{b}$: $T_{b}$=0.35 $\Gamma_{10}$(dot curves), $T_{b}$=0.50 $\Gamma_{10}$(dash curves), $T_{b}$=0.65 $\Gamma_{10}$(solid curves), $T_{a}$=0.24 $\Gamma_{10}$, and the other parameters are the same to Fig.3.}
\end{figure}\label{Fig.4}

In order to visualize the effect of inter-dot tunneling $T_{a}$ and $T_{b}$ on the left-handedness and absorption, Fig.3 and Fig.4 show $Re[\epsilon_{r}]$, $Re[\mu_{r}]$, $Re[n], $ $Im[n]$ of the TQD system versus the inter-dot tunneling $T_{a}$ and $T_{b}$ while keep the incoherent pumping field the same. As expected the left-handed behavior is observed in Fig.3 and Fig.4. In Fig.3 the pumping rate of the incoherent pumping field is set $\Gamma$=0.2$\Gamma_{10}$ ,and the tunneling $T_{a}$ is set $T_{a}$=0.23 $\Gamma_{10}$(dot curves), $T_{a}$=0.235 $\Gamma_{10}$(dash curves) and $T_{a}$=0.24 $\Gamma_{10}$(solid curves) which can be varied by the external gate voltage. The real parts of the the electric permittivity $Re[\epsilon_{r}]$, magnetic permeability $Re[\mu_{r}]$ and relative refractive index $Re[n]$ of the TQD system show the gradual increasing negative response via the increasing tunneling $T_{a}$. In the curves of $Im[n]$, there're three zero absorption windows, and two of them symmetrically distribute on both sides of the resonant dot when the tunneling $T_{a}$ is varied strongly.

As noted in Fig.4, the TQD system shows left-handedness with the negative $Re[\epsilon_{r}]$, $Re[\mu_{r}]$ and relative refractive index $Re[n]$ when the tunneling $T_{b}$ is varied by $T_{b}$=0.35 $\Gamma_{10}$(dot curves), $T_{b}$=0.50 $\Gamma_{10}$(dash curves), $T_{b}$=0.65 $\Gamma_{10}$(solid curves). While the three zero-absorption windows with two of them symmetrically distributing on both sides of the resonant dot still appear in the TQD system via the $Im[n]$ curves. The peculiar negative behaviors with zero-absorption of this TQD system interest us most, because our proposed scheme is significantly different from the photonic resonant media through the gaseous atomic system, and because the zero-absorption property of the left-handed TQD system may be applied to overcome the negative refractive materials with large amount of absorption. What's more, the manipulation via the external gate voltage and light field is a development technique for all optical communication\cite{36} and optoelectronic devices in the next generation.

\section{Conclusion}

In conclusion, the left-handedness with triple zero-absorption windows are demonstrated theoretically in a TQD system.
With the typical parameters of a GaAs/AlGa As heterostructure, the strong response of simultaneous negative relative electric permittivity and magnetic permeability were observed when the increasing incoherent pumping field and two inter-dot tunnelings between QD1 and QD2, QD2 and QD3
were applied to the TQD system. Then the left-handedness is achieved by the tuned pumping field and two inter-dot tunnelings. What's more, three zero absorption windows emerge in the left-handed frequency bands during the adjustable incoherent pumping field and inter-dot tunneling coupling the TQD system. The left-handedness with triple transparency windows in the GaAs/AlGaAs heterostructure may overcome not only the challenge in photonic resonant scheme for left-handedness, but also the main application limitation of the negative refractive materials with large amount of absorption. The proposed scheme is important because the flexible design and several controllable parameters can realize easily in a multi-dot system. Not only that, but light controlled by light or by electron tunneling are essential for all optical communication and optoelectronic devices in the next generation.


\begin{thebibliography}{0}%
\makeatletter
\providecommand \@ifxundefined [1]{%
 \@ifx{#1\undefined}
}%
\providecommand \@ifnum [1]{%
 \ifnum #1\expandafter \@firstoftwo
 \else \expandafter \@secondoftwo
 \fi
}%
\providecommand \@ifx [1]{%
 \ifx #1\expandafter \@firstoftwo
 \else \expandafter \@secondoftwo
 \fi
}%
\providecommand \natexlab [1]{#1}%
\providecommand \enquote  [1]{``#1''}%
\providecommand \bibnamefont  [1]{#1}%
\providecommand \bibfnamefont [1]{#1}%
\providecommand \citenamefont [1]{#1}%
\providecommand \href@noop [0]{\@secondoftwo}%
\providecommand \href [0]{\begingroup \@sanitize@url \@href}%
\providecommand \@href[1]{\@@startlink{#1}\@@href}%
\providecommand \@@href[1]{\endgroup#1\@@endlink}%
\providecommand \@sanitize@url [0]{\catcode `\\12\catcode `\$12\catcode
  `\&12\catcode `\#12\catcode `\^12\catcode `\_12\catcode `\%12\relax}%
\providecommand \@@startlink[1]{}%
\providecommand \@@endlink[0]{}%
\providecommand \url  [0]{\begingroup\@sanitize@url \@url }%
\providecommand \@url [1]{\endgroup\@href {#1}{\urlprefix }}%
\providecommand \urlprefix  [0]{URL }%
\providecommand \Eprint [0]{\href }%
\providecommand \doibase [0]{http://dx.doi.org/}%
\providecommand \selectlanguage [0]{\@gobble}%
\providecommand \bibinfo  [0]{\@secondoftwo}%
\providecommand \bibfield  [0]{\@secondoftwo}%
\providecommand \translation [1]{[#1]}%
\providecommand \BibitemOpen [0]{}%
\providecommand \bibitemStop [0]{}%
\providecommand \bibitemNoStop [0]{.\EOS\space}%
\providecommand \EOS [0]{\spacefactor3000\relax}%
\providecommand \BibitemShut  [1]{\csname bibitem#1\endcsname}%
\let\auto@bib@innerbib\@empty
%</preamble>
\end{thebibliography}%


\section*{References}
\begin{thebibliography}{ }
\bibitem{1}V. G. Veselago,``The electrodynamics of substances with simultaneously negative values of $\epsilon$ and $\mu$,''{\it Sov. Phys. Usp.} \textbf{10,} 509 (1968).
\bibitem{2}M. W. Feise, P. J. Bevelacqua and John B. Schneider,``Effects of surface waves on the behavior of perfect lenses,''{\it Phys. Rev. B} \textbf{66,} 035113 (2002)
\bibitem{3}K. Aydin, I. Bulu and E. Ozbay,``Subwavelength resolution with a negative-index metamaterial superlens,''{\it Appl. Phys. Lett. } \textbf{90,} 254102 (2007)
\bibitem{4}J. B. Pendry, A. J. Holden, W. J. Stewart, I. Youngs, ``Extremely low frequency plasmons in metallic mesostructures,''{\it Phys. Rev. Lett.} {\bf 76, } 4773(1996).
\bibitem{5}J. B. Pendry et al., ``Magnetism from conductors and enhanced nonlinear phenomena,''{\it IEEE Trans. Microwave Theory Tech.} {\bf 47, } 2075(1999).
\bibitem{6}R. A. Shelby, D. R. Smith, S. Schultz, ``Experimental veriTcation of a negative index of refraction,''{\it Science} {\bf 292, } 77 (2001).
\bibitem{7}E. Cubukcu, K. Aydin, E. Ozbay, S. Foteinopoulou C. M. Soukoulis,``Electromagnetic waves: Negative refraction by photonic crystals,''  {\it Nature} \textbf{423,} 604 (2003)
\bibitem{8}M. $\ddot{O}$. Oktel and $\ddot{O}$. E. M$\ddot{u}$stecapl$\breve{g}$lu, ``Electromagnetically induced left-handedness in a dense gas of three-level atoms,''{\it Phys. Rev. A} \textbf{ 70,} 053806 (2004 )
\bibitem{9}J. B. Pendry, ``A chiral route to negative refraction,'' {\it Science} \textbf{ 306,} 1353 (2004).
\bibitem{10}S. C. Zhao, Q. X. Wu and A. L. Gong,``Algebraic analysis of electromagnetic chirality-inducednegative refractive index in a four-level atomic system,''{\it Eur. Phys. J. D} \textbf{ 67,} 28 (2013).
\bibitem{11}S. He, Z. Ruan, L. Chen and J. Shen,`` Focusing properties of a photonic crystal slab with negative refraction,''{\it Phys. Rev. B } \textbf{70,} 115113 (2004).
\bibitem{12}J. Q. Shen, Z. C. Ruan and S. He,``Influence of the signal light on the transient optical properties of a four-level EIT medium,'' {\it Phys. Lett. A} \textbf{330,} 487 (2004).
\bibitem{13}Q. Thommen and P. Mandel,``Electromagnetically induced left handedness in optically excited four-level atomic media,''{\it Phys. Rev. Lett. } \textbf{96,} 053601 (2006).
\bibitem{14}S. C. Zhao, Z. D. Liu, J. Zheng and Z.Q. Zhang,``Negative refraction with absorption suppressed by EIT in a left-handed atomic system.,''{\it Sci. Chin. Physics, Mechanics Astronomy} \textbf{55,} 213 (2012).
\bibitem{15}F. R. Waugh, M. J. Berry, D. J. Mar, R. M. Westervelt, K. L. Campman, and A. C. Gossard, ``Single-electron charging in double and triple quantum dots with tunable coupling,''{\it Phys. Rev. Lett. } \textbf{75,} 705 (1995).
\bibitem{16}A. D. Greentree, J. H. Cole, A. R. Hamilton, and L. C. L.Hollenberg, ``Coherent electronic transfer in quantum dot systems using adiabatic passage," {\it Phys. Rev. B} \textbf{ 70,} 235317 (2004).
\bibitem{17}C. Kloeffel and D. Loss,``Prospects for spin-based quantum computing in quantum dots," {\it Annu. Rev. Condens. Matter Phys.} \textbf{4,} 51 (2013)
\bibitem{18}L. Gaudreau, A. Kam, G. Granger, S. A. Studenikin, P. Zawadzki, and A. S. Sachrajda,``A tunable few electron triple quantum dot," {\it Appl. Phys. Lett.} \textbf{ 95,} 193101 (2009).
\bibitem{19}L. Gaudreau, G. Granger, A. Kam, G. C. Aers, S. A. Studenikin, P. Sawadzki, M. Pioro-Ladri$\acute{`e}$re, Z. R. Wasilewski, and A. S.
    Sachrajda, ``Coherent control of three-spin states in a triple quantum dot," {\it Nat. Phys.} \textbf{ 8,} 54 (2012).
\bibitem{20}J. Villavicencio, I. Maldonado, E. Cota and G. Platero,``Spin-orbit effects in a triple quantum dot shuttle," {\it Phys. Rev. B} \textbf{ 88,} 245305 (2013).
\bibitem{21}J. Medford, J. Beil, J. M. Taylor, S. D. Bartlett, A. C. Doherty, E. I. Rashba, D. P. Divincenzo, H. Lu, A. C. Gossard, and
    C. M. Marcus, ``Self-consistent measurement and state tomography of an exchange-only spin qubit," {\it Nat. Nanotechnol.} \textbf{8,} 654 (2013).
\bibitem{22}M. Busl, G. Granger, L. Gaudreau, R. S$\acute{a}$nchez, A. Kam, M. PioroLadri$\acute{e}$re, S. A. Studenikin, P. Zawadzki, Z. R. Wasilewski, A. S. Sachrajda, and G. Platero,``Bipolar spin blockade and coherent state superpositions in a triple quantum dot," {\it Nat. Nanotechnol.} \textbf{ 8,} 261 (2013).
\bibitem{23}C. Y. Hsieh, Y. P. Shim, and P. Hawrylak, ``Theory of electronic properties and quantum spin blockade in a gated linear triple quantum dot with one electron spin each," {\it Phys. Rev. B} \textbf{ 85,} 085309 (2012).
\bibitem{24}M. H. You, Z. G. Li, X. Gao, X. D. Liu, Y. Deng, G. J. Liu, L. Li, Z. P. Wei and X. H. Wang, ``Long wavelength strain engineered InAs multi-layer stacks quantum dots laser diode on GaAs substrate," {\it Laser Phys.} \textbf{22,} 1673 (2012).
\bibitem{25}D. Birkedal, K. Leosson and J. M. Hvam, `` Long lived coherence in self-assembled quantum dots," {\it Phys. Rev. Lett.} \textbf{ 87,} 227401 (2001).
\bibitem{26}J. Bardeen, ``Tunneling from a many-particle point of view,"  {\it Appl. Phys. Lett.} \textbf{6,} 57 (1961)
\bibitem{27}H. J. Reittu, ``Fermi's golden rule and Bardeen's tunneling theory,"  {\it Am. J. Phys.} \textbf{63,} 940 (1995).
\bibitem{28}O. Kocharovskaya, Y. Rostovtsev and M. O. Scully, ``Stopping light via hot atoms," {\it Phys. Rev. Lett.}  \textbf{ 86,} 628 (2001)
\bibitem{29}S. C. Zhao, S. Y. Zhang, and Y. Y. Xu, ``Large and tunable negative refractive index via electromagnetically induced chirality in a
          semiconductor quantum well nanostructure," {\it JETP Lett.}  \textbf{ 100,} 385 (2014).
\bibitem{30}C. S. Zhao, D. Z. Liu, ``Left handness in a four-level atomic system," {\it Int. J. Quant. Inf.}  \textbf{ 7,} 747 (2009).
\bibitem{31}C. S. Zhao, D. Z. Liu, Q. X. Wu, ``Negative refraction without absorption via both coherent and incoherent fields in a
         four-level left-handed atomic system," {\it Opt. Comms.} \textbf{ 283,} 3301 (2010).
\bibitem{32}F. R. Waugh, M. J. Berry, D. J. Mar and R. M. Westervelt, ``Single-electron charging in double and triple quantum dots with tunable coupling," {\it Phys. Rev. Lett.}  \textbf{75,} 705 (1995).
\bibitem{33}K. H. Yoo, L. R. Ram Mohan and D. F. Nelson,``Effect of nonparabolicity in $GaAs/Ga_{1-x}Al_{x}$ as semiconductor quantum wells," {\it Phys. Rev. B} \textbf{ 39,} 12808 (1989).
\bibitem{34}F. L. Li, A. P. Fang, and M. Wang,``Electromagnetic chirality-induced negative refraction via atomic coherenc," {\it J. Phys. B: At Mol. Opt. Phys.} \textbf{ 42,} 199505 (2009).
\bibitem{35}M. R. Mehmannavaz and H. Sattari,``Electrical and optical control of optical gain in a coupled triple quantum dot system operating in telecommunication window," {\it Laser Phys. } \textbf{ 24,} 125201 (2014).
\bibitem{36}H. S. Borges, L. Sanz, J. M. Villas-Boas and A. M. Alcalde, ``Quantum interference and control of the optical response in quantum dot molecules," {\it Appl. Phys. Lett.} \textbf{103,} 222101 (2013).
\end{thebibliography}
\end{document}